\documentclass[ISTS]{tjsass} 

\usepackage{graphicx}
\RequirePackage{multicol}
\RequirePackage{amsmath}
\RequirePackage[varg]{txfonts}
\RequirePackage{bm}
\RequirePackage{array}
\RequirePackage{color}
\RequirePackage{tjsasscite}
\usepackage[hang]{footmisc}
\newcommand{\bhline}[1]{\noalign{\hrule height #1}}
\usepackage{makecell} 
\usepackage{subfigure}

\pubyear{2025}
\bookvolume{17}
\bookissue{20}
\titleheadertrue

\title{Improving Spatio-Temporal Accuracy of the Stochastic Particle Fokker-Planck Model}
\subtitle{}
\author{\NAME{Joonbeom}{Kim}\thanksNum{1)}\CorresAuthor{kjb9910@kaist.ac.kr} and \NAME{Eunji}{Jun}\thanksNum{1)}}
\thanksOrg{Department of Aerospace Engineering, Korea Advanced Institute of Science and Technology, Daejeon, South Korea} 

\begin{abstract}

Accurate prediction of rarefied gas flows is important for space vehicle design, particularly in rarefied regimes where the Navier-Stokes equations are no more valid. While the direct simulation Monte Carlo (DSMC) method acts as a numerical solver for rarefied gas flows, it becomes inefficient when dealing with near-continuum regimes. The Fokker-Planck (FP) model improves computational efficiency by approximating particle collisions as a drift-diffusion process. The FP model has been extended to handle diatomic gases, such as the Fokker-Planck-Master (FPM) model. The FPM model's first-order accuracy in both time and space limits computational efficiency gains. This study proposes a unified stochastic particle FPM (USP-FPM) model that achieves second-order spatio-temporal accuracy for diatomic gases. Temporal accuracy is improved by introducing second-order energy relaxation into the USP-FP method. Spatial accuracy is improved by employing a polynomial reconstruction method for macroscopic properties. The USP-FPM model is validated through two numerical simulations: relaxation to thermal equilibrium in a homogeneous flow and hypersonic flow over a vertical plate. The results demonstrate that the USP-FPM model shows good agreement with DSMC results and significantly reduces computational cost by enabling larger cell sizes and time steps.

\end{abstract}
\keywords{Rarefied gas flows, Boltzmann equation, Direct Simulation Monte Carlo, Fokker-Planck method}
\begin{document}
\maketitle


\section*{Nomenclature}

\vbox{\noindent\setlength{\tabcolsep}{0mm}%
\begin{tabular}{p{25mm}cl} %

\hfil$A$\hfil & :\hspace{4mm} & Drift coefficient \\
\hfil$C$\hfil & :\hspace{4mm} & Thermal velocity of particle \\
\hfil$c$\hfil & :\hspace{4mm} & Velocity of particle \\
\hfil$c_v$\hfil & :\hspace{4mm} & Specific heat capacity \\
\hfil$c_\omega$\hfil & :\hspace{4mm} & Viscosity index \\
\hfil$D$\hfil & :\hspace{4mm} & Diffusion coefficient \\
\hfil$d_{ij}$\hfil & :\hspace{4mm} & components of the matrix \\
                   &   & satisfying $D_{\mathrm{trn},ij} = \sum_{k} d_{ik} d_{kj}$ \\
\hfil$e$\hfil & :\hspace{4mm} & Specific energy \\
\hfil$\mathcal{F}$\hfil & :\hspace{4mm} & Mass density function \\
\hfil$f$\hfil & :\hspace{4mm} & Probability density function \\
\hfil$G$\hfil & :\hspace{4mm} & Gaussian random number \\
\hfil$I$\hfil & :\hspace{4mm} & Vibrational energy level \\
\hfil$\mathrm{int}(\cdot)$\hfil & :\hspace{4mm} & Integer part \\ 
\hfil$k_B$\hfil & :\hspace{4mm} & Boltzmann constant \\
\hfil$m$\hfil & :\hspace{4mm} & Molecular mass \\
\hfil$\mathrm{N_{p}}$\hfil & :\hspace{4mm} & Number of particles \\
\hfil$\mathrm{N_{sampling}}$\hfil & :\hspace{4mm} & Number of time step for sampling \\
\hfil$\mathrm{N_{steady}}$\hfil & :\hspace{4mm} & Number of time step to reach  \\
                          &               & steady state \\
\hfil$n$\hfil & :\hspace{4mm} & Number density \\
\hfil$P_\mathrm{BD}$\hfil & :\hspace{4mm} & Random sample from the  \\
                          &               & binomial distribution \\
\hfil$P_\mathrm{NBD}$\hfil & :\hspace{4mm} & Random sample from the  \\
                          &                & negative binomial distribution \\
\hfil$p$\hfil & :\hspace{4mm} & Pressure \\
\hfil$R$\hfil & :\hspace{4mm} & Specific gas constant \\
\hfil$T$\hfil & :\hspace{4mm} & Temperature \\
\hfil$\tilde{T}$\hfil & :\hspace{4mm} & Auxiliary temperature \\
\hfil$t$\hfil & :\hspace{4mm} & Time \\
\hfil$U$\hfil & :\hspace{4mm} & Bulk velocity \\
\end{tabular}}

\vbox{\noindent\setlength{\tabcolsep}{0mm}%
\begin{tabular}{p{25mm}cl} %
\hfil$\mathrm{Unif}(0,1)$\hfil & :\hspace{4mm} & Random variable uniformly distributed \\
                               &               & on the interval (0,1) \\
\hfil$W$\hfil & :\hspace{4mm} & Standard Wiener process \\
\hfil$x$\hfil & :\hspace{4mm} & Position of particle \\
\hfil$Z$\hfil & :\hspace{4mm} & Collision number \\
\hfil$\alpha$\hfil & :\hspace{4mm} & Time integration rate \\
\hfil$\gamma$\hfil & :\hspace{4mm} & Energy transfer coefficient \\
\hfil$\Delta t$\hfil & :\hspace{4mm} & Time step \\
\hfil$\Delta x$\hfil & :\hspace{4mm} & Cell size \\
\hfil$\delta$\hfil & :\hspace{4mm} & Kronecker delta \\
\hfil$\varepsilon$\hfil & :\hspace{4mm} & Energy of particle \\
\hfil$\Theta$\hfil & :\hspace{4mm} & Characteristic temperature \\
\hfil$\lambda$\hfil & :\hspace{4mm} & Eigenvalue of temperature tensor\\
\hfil$\mu$\hfil & :\hspace{4mm} & Viscosity \\
\hfil$\nu$\hfil & :\hspace{4mm} & Parameter for the Prandtl number \\
\hfil$\xi$\hfil & :\hspace{4mm} & Degrees of freedom \\
\hfil$\Pi$\hfil & :\hspace{4mm} & Temperature tensor \\
\hfil$\rho$\hfil & :\hspace{4mm} & Density \\
\hfil$\tau$\hfil & :\hspace{4mm} & Relaxation time \\
\hfil$\tau_c$\hfil & :\hspace{4mm} & Mean collision time \\
\hfil$\phi$\hfil & :\hspace{4mm} & Macroscopic property \\
\hfil$\chi$\hfil & :\hspace{4mm} & Scaling factor \\
\hfil$\overline{\phi}$\hfil & :\hspace{4mm} & Cell-averaged macroscopic property \\
\hfil$\omega$\hfil & :\hspace{4mm} & Transition rate coefficient \\
\end{tabular}}

\noindent{Subscripts}

\vbox{\noindent\setlength{\tabcolsep}{0mm}%
\begin{tabular}{p{25mm}cl}

\hfil$0$\hfil & :\hspace{4mm} & initial \\
\hfil$\mathrm{ref}$\hfil & :\hspace{4mm} & reference \\
\hfil$\mathrm{rot}$\hfil & :\hspace{4mm} & rotational mode \\
\hfil$\mathrm{trn}$\hfil & :\hspace{4mm} & translational mode \\
\hfil$\mathrm{vib}$\hfil & :\hspace{4mm} & vibrational mode \\
\hfil$\mathrm{wall}$\hfil & :\hspace{4mm} & wall \\
\hfil$\Delta$\hfil & :\hspace{4mm} & difference \\
\hfil$\mathrm{\infty}$\hfil & :\hspace{4mm} & freestream \\
\end{tabular}}

\noindent{Superscripts}

\vbox{\noindent\setlength{\tabcolsep}{0mm}%
\begin{tabular}{p{25mm}cl}

\hfil$n$\hfil & :\hspace{4mm} & value at time $t$ \\
\hfil$n+1$\hfil & :\hspace{4mm} & value at time $t+\Delta t$ \\
\hfil$\mathrm{rel}$\hfil & :\hspace{4mm} & relaxation \\
\hfil$*$\hfil & :\hspace{4mm} & post-collision value before correction \\
\end{tabular}}

\section{\label{sec:intro}Introduction}


As the new space age emerges, accurate prediction of flow fields becomes important for designing safe and efficient space vehicles. The traditional Navier-Stokes equations become inadequate in the upper atmosphere, where the flow transitions to a rarefied regime. For such regimes, the Boltzmann equation more accurately describes gas dynamics.\cite{Book-Bird94-DSMC, KIM23-CompareCFDKinetic} A widely used numerical approach to solve the Boltzmann equation is the direct simulation Monte Carlo (DSMC) method. While effective across all flow regimes, the DSMC method becomes computationally expensive in near-continuum flows.


To deal with the limitations, several alternatives have been developed such as kinetic modeling and DSMC-hybrid methods.\cite{Jun11-Particle-conf, Jun13-Particle-Conf, Jun18-LD-DSMC, Jun18-FPDSMC-cylinder, Jun19-FPDSMC-diatomic, Kim24-FPDSMC, Yai23-SBGK, Park24-BGK-PDF} Among these, Jenny et al. pioneered the stochastic particle Fokker–Planck (FP) model, which approximates particle collisions as a drift–diffusion process.\cite{Jenny10_linearFP} The linear FP model suffers from physical inaccuracies, such as predicting an incorrect Prandtl number for monatomic gases. To improve physical fidelity, several advanced FP models have been proposed, including the cubic Fokker–Planck (Cubic-FP) and ellipsoidal statistical Fokker–Planck (ES-FP) models.\cite{Gorji11_cubicFP, Mathiaud15_ESFP, Gorji21_QEFP, Kim22-FPNonlinear, Jun19-CompareFP, Kim24-AssessFP-Conf} Moreover, the FP model has been extended to handle polyatomic gases and mixtures.\cite{Gorji13_CubicFP-diatomic, Jun18-CubicFP-Slit, Mathiaud17-ESFP-polyatomic, Hepp20-CubicFP-diatomic, Kim24-FPM, Gorji12-CubicFP-monamix, Hepp20-CubicFP-HSmix, Hepp20-CubicFP-VHSmix, Kim25-ESFP-monamix, Kim25-ESFPM-dimix}


Although there has been progress in physical fidelity and applicability, relatively few studies have focused on the spatio-temporal accuracy of FP models. In conventional FP models, particle velocities are updated using cell-averaged macroscopic properties, which neglect spatial gradients and lead to first-order spatial accuracy.\cite{Kim23-AssessmentFP} Since particle transport and collisions are decoupled, FP models also exhibit first-order temporal accuracy. This leads to constraints on cell size and time step, limiting computational efficiency. Recently, Kim et al. proposed the unified stochastic particle FP (USP-FP) method, which achieves second-order temporal accuracy.\cite{Kim24-USPFP} Separately, Cui et al. developed the multiscale stochastic particle (MSP) method, providing a mathematical proof of second-order temporal accuracy.\cite{Cui25-MSP} Kim et al. improves spatial accuracy of the FP model by employing the polynomial reconstruction method.\cite{Kim24-poly} These studies have been limited to monatomic gases, and the extension to diatomic gases remains unexplored.


This study develops a unified stochastic particle Fokker-Planck-Master (USP-FPM) model that achieves second-order spatio-temporal accuracy for diatomic gases. Temporal accuracy is enhanced by extending the USP-FP method to incorporate energy relaxation. Spatial accuracy is improved through a polynomial reconstruction method for interpolating macroscopic properties. Section \ref{sec:review} reviews the FPM model for diatomic gases. Section \ref{sec:usp-fpm} presents improvements in spatial and temporal accuracy of the FPM model. Section \ref{sec:implementation} describes numerical implementations applied in the analysis. Section \ref{sec:results} provides numerical results for two test cases: relaxation to thermal equilibrium in a homogeneous flow and hypersonic flow over a vertical plate. Finally, conclusions are drawn in Section \ref{sec:conclusions}.

\section{\label{sec:review}Review of stochastic particle Fokker–Planck–Master model}


In kinetic theory, the probability distribution function (PDF), denoted as $f(t, \boldsymbol{x}, \boldsymbol{c}, \varepsilon_{\mathrm{rot}}, \varepsilon_{\mathrm{vib}})$, describes the statistical behavior of diatomic gases.\cite{Book-Bird94-DSMC} Above room temperature, the rotational energy can be treated as continuous, whereas the vibrational energy remains quantized. Using the harmonic oscillator model, the vibrational energy is expressed as $\varepsilon_{\mathrm{vib}} = IR\Theta_{\mathrm{vib}}$. Each energy mode has its specific energy defined as:
\begin{equation}
    e_{\mathrm{trn}}(T) = \frac{3}{2}RT
\end{equation}
\begin{equation}
    e_{\mathrm{rot}}(T) = RT
\end{equation}
\begin{equation}
    e_{\mathrm{vib}}(T) = \frac{R\Theta_{\mathrm{vib}}}{\exp{(\Theta_{\mathrm{vib}}/T)}-1}
\end{equation}
The FPM model describes the time evolution of the PDF using a drift–diffusion–jump mechanism.\cite{Kim24-FPM} The FP equation models the evolution of the continuous velocity and rotational energy, whereas the master equation describes the discrete vibrational energy. The mass density function (MDF), defined as $\mathcal{F} = \rho f$, is used to express the FPM model in terms of mass density. The FPM model is expressed as:
\begin{multline}
    \frac{\partial \mathcal{F}}{\partial t} 
    + c_j\frac{\partial \mathcal{F}}{\partial c_j} 
    = - \frac{\partial}{\partial c_i} (A_{\mathrm{trn}, i} \, \mathcal{F})
    + \frac{\partial^2}{\partial c_i \partial c_j} (D_{\mathrm{trn}, ij} \, \mathcal{F})\\
     - \frac{\partial}{\partial \varepsilon_{\mathrm{rot}}} (A_{\mathrm{rot}} \, \mathcal{F})
    + \frac{\partial^2}{\partial \varepsilon_{\mathrm{rot}}} (D_{\mathrm{rot}} \, \mathcal{F})\\
     + \sum_{J=0}^{\infty} (\omega_{J,I} \, \mathcal{F}_J - \omega_{I,J} \, \mathcal{F}_I) .
\label{eq:01_FPM}
\end{multline}
Einstein notation is adopted for repeated indices throughout this paper. The model coefficients are designed to maintain consistency with the Navier–Stokes–Fourier equations in the continuum limit:
\begin{equation}
        A_{\mathrm{trn},i} = -\frac{C_i}{\tau} ,
\label{eq:02.1_Atr}
\end{equation}
\begin{equation}
    D_{\mathrm{trn},ij} = \frac{R T_{\mathrm{trn}}^{\mathrm{rel}}\, \delta_{ij} + \nu (\Pi_{ij} - R T_{\mathrm{trn}}\, \delta_{ij})}{\tau} ,
\label{eq:02.2_Dtr}
\end{equation}
\begin{equation}
    A_{\mathrm{rot}} = -\frac{2}{\tau} \Bigl(\varepsilon_{\mathrm{rot}} - RT_{\mathrm{rot}}^{\mathrm{rel}} \Bigr) ,
\label{eq:02.3_Arot}
\end{equation}
\begin{equation}
    D_{\mathrm{rot}} = \frac{2RT_{\mathrm{rot}}^{\mathrm{rel}} \, \varepsilon_{\mathrm{rot}}}{\tau} ,
\label{eq:02.4_Drot}
\end{equation}
\begin{equation}
    \omega_{I,J} = 
        \begin{cases}
            \frac{2 I}{\tau (1 - \exp{(-\Theta_{\mathrm{vib}}/T_{\mathrm{vib}}^{\mathrm{rel}})})} & \text{if } J = I - 1, \\[1em]
            \frac{2 (I+1) \exp{(-\Theta_{\mathrm{vib}}/T_{\mathrm{vib}}^{\mathrm{rel}})}}{\tau (1 - \exp{(-\Theta_{\mathrm{vib}}/T_{\mathrm{vib}}^{\mathrm{rel}})})} & \text{if } J = I + 1, \\[1em]
            0 & \text{otherwise},
        \end{cases}
\label{eq:02.5_omega}
\end{equation}
\begin{equation}
    \tau = \frac{2\mu (1-\nu)}{p} .
\label{eq:02.6_tau}
\end{equation}
The thermal velocity is defined as $C_i=c_i-U_i$, and the relaxation temperatures $T^{\mathrm{rel}}$ control the internal energy relaxation. The Prandtl number $(\mathrm{Pr})$ is given by:
\begin{equation}
    \mathrm{Pr} = \frac{3}{2(1-\nu)}.
\end{equation}

\section{\label{sec:usp-fpm}Unified stochastic particle Fokker–Planck–Master model for diatomic gas flows}

\subsection{\label{subsec:moment}Temporal Evolution}

\subsubsection{\label{subsubsec:evolution}Particle evolution scheme in the USP-FPM Model}


The evolution of a particle's position and velocity is described by the equivalent Langevin equations:\cite{Jenny10_linearFP}
\begin{equation}
    dx_i=c_idt ,
\label{eq:03_dx}
\end{equation}
\begin{equation}
    dC_i=A_{\mathrm{trn},i} \, dt+\sqrt{2}d_{ij} \, dW_{j}.
\label{eq:04_dc}
\end{equation}
Gorji's time integration scheme is commonly used to describe the time evolution of a particle's position and velocity.\cite{Gorji14-effect} A limitation of this scheme is the assumption that macroscopic properties remain constant during each time step, neglecting their temporal evolution. Kim et al. proposed the USP-FP method to achieve second-order temporal accuracy by reformulating the time integration scheme.\cite{Kim24-USPFP} To extend the USP-FP method to diatomic gases, internal energy relaxation processes must be considered. Based on the unified stochastic particle Bhatnagar-Gross-Krook (USP-BGK) model for polyatomic gases and the ellipsoidal statistical BGK (ES-BGK) model that incorporates the Landau–Teller equation, this study proposes a unified stochastic particle FPM (USP-FPM) model for diatomic gases, which achieves second-order accuracy. \cite{Fei22-USPBGK-poly, Mathiaud22-ESBGK} The particle velocity update scheme is formulated as follows:
\begin{equation}
    C_i^{n+1}=C_i^n \, \alpha + \sqrt{1 - \alpha^2} \, d_{\mathrm{trn},ij} \, G_j ,
\label{eq:dc}
\end{equation}
\begin{equation}
    \alpha=\Bigl( \frac{2\mu/p-\mathrm{Pr} \Delta t}{2\mu/p+\mathrm{Pr} \Delta t} \Bigr)^{1/3} ,
\label{eq:da}
\end{equation}
\begin{equation}
    \nu= \frac{1}{1-\alpha^2} \, \Bigl(\frac{2\mu/p - \Delta t}{2\mu/p + \Delta t} - \alpha^2 \Bigr) .
\label{eq:dnu}
\end{equation}
To reflect second-order energy relaxation, the translational relaxation temperature is redefined as follows:
\begin{multline}
    e_{\mathrm{trn}}(T_{\mathrm{trn}}^{\mathrm{rel}}) = e_{\mathrm{trn}}(T_{\mathrm{trn}})
    - \frac{\gamma_{\mathrm{trn-rot}}}{1-\alpha^2} (T_{\mathrm{trn}} - T_{\mathrm{rot}}) \\[0.5em]
    - \frac{\gamma_{\mathrm{trn-vib}}}{1-\alpha^2} (T_{\mathrm{trn}} - T_{\mathrm{vib}}),
\end{multline}
\begin{equation}
    \gamma_{\mathrm{trn-rot}} = 
    \frac
    {
        \frac{2\Delta t}{\Delta t + 2\tau_{\mathrm{rot}}}
        c_{v,\mathrm{trn}} \, c_{v,\mathrm{rot}}
    }
    {
        c_{v,\mathrm{trn}} 
        + \frac{\Delta t}{\Delta t + 2\tau_{\mathrm{rot}}} c_{v,\mathrm{rot}} 
        + \frac{\Delta t}{\Delta t + 2\tau_{\mathrm{vib}}} c_{v,\mathrm{vib}}(T_1)
    },
\end{equation}
\begin{equation}
    \gamma_{\mathrm{trn-vib}} = 
    \frac
        {
            \frac{2\Delta t}{\Delta t + 2\tau_{\mathrm{vib}}} 
            c_{v,\mathrm{trn}} \, c_{v,\mathrm{vib}}(T_1)
        }
        {
            c_{v,\mathrm{trn}} 
            + \frac{\Delta t}{\Delta t + 2\tau_{\mathrm{rot}}} c_{v,\mathrm{rot}} 
            + \frac{\Delta t}{\Delta t + 2\tau_{\mathrm{vib}}} c_{v,\mathrm{vib}}(T_1)
        },
\end{equation}
The translational and rotational specific heat capacities are treat as constant, given by $c_{v,\mathrm{trn}}=3R/2$ and $c_{v,\mathrm{rot}}=R$, respectively. In contrast, the vibrational specific heat capacity varies with temperature. Here, it is approximated using the mean value theorem between the translational temperature $T_{\mathrm{trn}}$ and an auxiliary vibrational temperature $\tilde{T}_{\mathrm{vib}}$:
\begin{equation}
    c_{v,\mathrm{vib}}(T_1)
    = \frac{e_{\mathrm{vib}}(T_{\mathrm{trn}}) - e_{\mathrm{vib}}(\tilde{T}_{\mathrm{vib}}) }{T_{\mathrm{trn}} - \tilde{T}_{\mathrm{vib}}} ,
\end{equation}
\begin{equation}
    e_{\mathrm{vib}}(\tilde{T}_{\mathrm{vib}})
     = e_{\mathrm{vib}}(T_{\mathrm{vib}})
     - \frac{\Delta t}{2\tau_{\mathrm{vib}}}
        (e_{\mathrm{vib}}(T_{\mathrm{trn}}) - e_{\mathrm{vib}} (T_{\mathrm{vib}}) ) .
\end{equation}


For the rotational energy, the Milstein scheme is employed to ensure the positivity of the rotational energy:\cite{Kim24-FPM}
\begin{equation}
    \varepsilon_{\mathrm{rot}}^{n+1}
    =
    \frac{RT_{\mathrm{rot}}^{\mathrm{rel}}}{2} (1 - \alpha^2 )
    \,+\, 
    \biggl(
        \sqrt{\varepsilon_{\mathrm{rot}}^{n}} \, \alpha
    + \sqrt{\frac{RT_{\mathrm{rot}}^{\mathrm{rel}}}{2} (1 - \alpha^2 )} \, G
    \biggr)^2 .
\end{equation}
The rotational relaxation temperature is redefined as:
\begin{multline}
    e_{\mathrm{rot}}(T_{\mathrm{rot}}^{\mathrm{rel}}) = e_{\mathrm{rot}}(T_{\mathrm{rot}})
    + \frac{\gamma_{\mathrm{trn-rot}}}{1-\alpha^2} (T_{\mathrm{trn}} - T_{\mathrm{rot}}) \\[0.5em]
    - \frac{\gamma_{\mathrm{rot-vib}}}{1-\alpha^2} (T_{\mathrm{rot}} - T_{\mathrm{vib}}) ,
\end{multline}
\begin{equation}
    \gamma_{\mathrm{rot-vib}} = 
    \frac
    {
        \frac{2\Delta t}{\Delta t + 2\tau_{\mathrm{rot}}}
        \frac{\Delta t}{\Delta t + 2\tau_{\mathrm{vib}}}
        c_{v,\mathrm{rot}} \, c_{v,\mathrm{vib}}(T_1)
    }
    {
        c_{v,\mathrm{trn}} 
            + \frac{\Delta t}{\Delta t + 2\tau_{\mathrm{rot}}} c_{v,\mathrm{rot}} 
            + \frac{\Delta t}{\Delta t + 2\tau_{\mathrm{vib}}} c_{v,\mathrm{vib}}(T_1)
    }.
\end{equation}


The temporal evolution of vibrational energy is approximated using a modified tau-leaping method applied at each discrete time step. \cite{Gillespie00-tau-leap, Chatterjee05-modified-tau-leap} In this method, vibrational excitations are sampled from a negative binomial distribution, while de-excitation are drawn from a binomial distribution. Accordingly, the stochastic evolution of the vibrational energy level is described as follows:
\begin{multline}
    I^{n+1}
    =
    I^{n}
    \,+\, 
    P_{NBD} 
        \Big(
            (I^n+1) 
            \,;\, 
            \frac{1}{
                1 
                + \frac{\exp{(-\Theta_{\mathrm{vib}}/T_{\mathrm{vib}}^{\mathrm{rel}})}}{1-\exp{(-\Theta_{\mathrm{vib}}/T_{\mathrm{vib}}^{\mathrm{rel}})}}  
                (1-\alpha^2)
                }
        \Big) \\[1em]
    \,-\, 
    P_{BD}
    \Big(
        I^n 
        \,;\,
        \frac{1}{1-\exp{(-\Theta_{\mathrm{vib}}/T_{\mathrm{vib}}^{\mathrm{rel}})}} 
        (1-\alpha^2)
    \Big) .
\end{multline}
The vibrational relaxation temperature is redefined as:
\begin{multline}
    e_{\mathrm{vib}}(T_{\mathrm{vib}}^{\mathrm{rel}}) = e_{\mathrm{vib}}(T_{\mathrm{vib}})
    + \frac{\gamma_{\mathrm{trn-vib}}}{1-\alpha^2} (T_{\mathrm{trn}} - T_{\mathrm{vib}}) \\[0.5em]
    + \frac{\gamma_{\mathrm{rot-vib}}}{1-\alpha^2} (T_{\mathrm{rot}} - T_{\mathrm{vib}}) .
\end{multline}

\subsubsection{\label{subsubsec:positivity}Discussion on positivity of diffusion tensor}


The translational diffusion tensor must remain positive definite to ensure physical consistency.\cite{Kim24-FPM} Since the USP-FPM and FPM models use the same definition of the translational diffusion tensor, the associated positivity condition is identical and is given by:
\begin{equation}
    -\frac{RT_{\mathrm{trn}}^{\mathrm{rel}}}{\lambda_{\mathrm{max}}-RT_{\mathrm{trn}}} 
    <\nu<
    \frac{RT_{\mathrm{trn}}^{\mathrm{rel}}}{RT_{\mathrm{trn}} - \lambda_{\mathrm{min}}} 
\end{equation}
Depending on the flow condition, the reference value $\nu_{\mathrm{ref}}$ calculated from Eq. (\ref{eq:dnu}) may violate the positivity condition. To ensure positivity, $\nu$ is selected according to the following criterion:
\begin{equation}
    \nu = 
    \begin{cases}
        \max{\bigl( 
            \nu_{\mathrm{ref}}, 
            -\frac{RT_{\mathrm{trn}}^{\mathrm{rel}}}{\lambda_{\mathrm{max}}-RT_{\mathrm{trn}}} 
        \bigr)} & \text{if } \nu_{\mathrm{ref}} \leq 0 , \\[1em]
        \max{\bigl( 
            \nu_{\mathrm{ref}}, 
            \frac{RT_{\mathrm{trn}}^{\mathrm{rel}}}{RT_{\mathrm{trn}} - \lambda_{\mathrm{min}}}  
        \bigr)} & \text{otherwise}.
    \end{cases}
\end{equation}
Varying the parameter $\nu$ affects both the Prandtl number and the order of accuracy. Although an incorrect Prandtl number is an inherent limitation of the FPM model, the order of accuracy can be preserved by redefining $\alpha$ as:
\begin{equation}
    \alpha^2 = \frac{ 1 }{\nu - 1} \Bigl( \nu - \frac{2\mu/p - \Delta t}{2\mu/p + \Delta t} \Bigr)
\end{equation}
From this equation, two expressions for $\alpha$ can be derived:
\begin{equation}
    \alpha_1 = \sqrt{\alpha^2} 
    \text{ or } 
    \alpha_2 = -\sqrt{\alpha^2} .
\end{equation}
Each candidate $\alpha$ corresponds to a distinct Prandtl number, given by: 
\begin{equation}
    \mathrm{Pr}_1 = \frac{2\mu/p \, (1 - \alpha_1^3)}{\Delta t \, (1 + \alpha_1^3)} 
    \text{ or } 
    \mathrm{Pr}_2 = \frac{2\mu/p \, (1 - \alpha_2^3)}{\Delta t \, (1 + \alpha_2^3)}  ,
\end{equation}
Between the two candidates, the value of $\alpha$ that minimizes the deviation from the target Prandtl number is chosen.

\subsection{\label{subsec:spatial}Spatial Reconstruction}


\begin{figure}[!t]%
\centering
\includegraphics[width=75mm,clip]{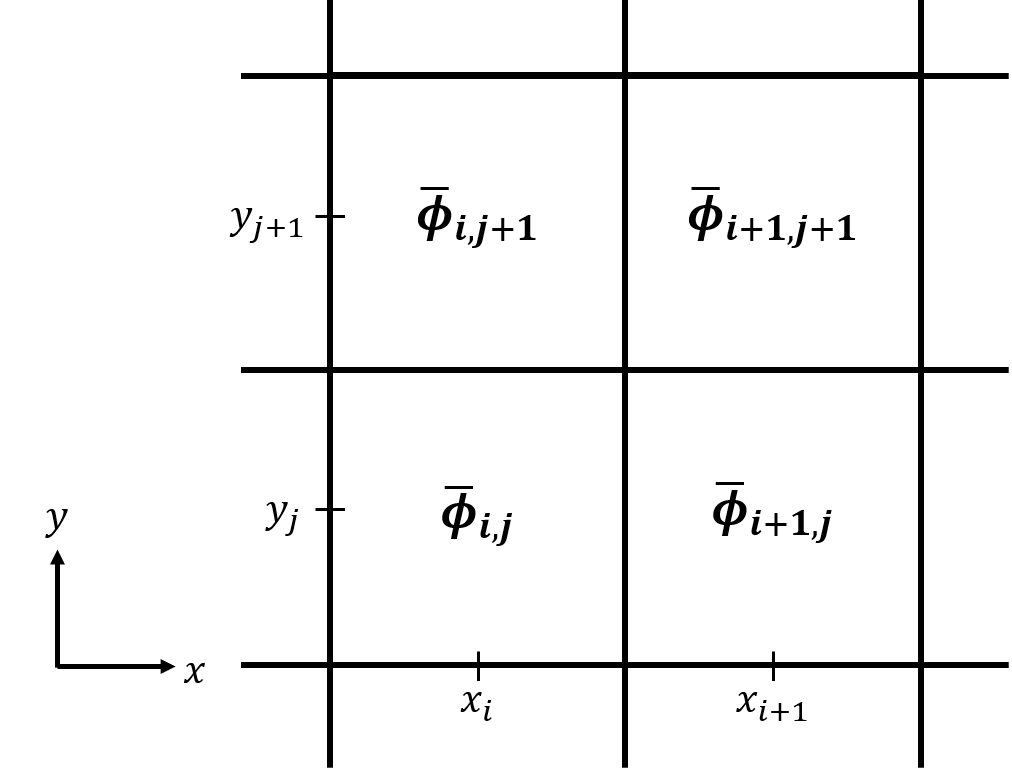}
\caption{A schematic of a uniform grid used for linear polynomial reconstruction.}%
  \label{fig:poly_uniform}%
\end{figure}%

The FPM model exhibits first-order spatial accuracy due to the use of cell-averaged macroscopic properties.\cite{Kim23-AssessmentFP} To achieve second-order spatial accuracy, the gradient of macroscopic properties within each cell needs to be considered. A first-order polynomial reconstruction method is employed to capture gradients, thereby improving spatial accuracy.\cite{Kim24-poly} For each macroscopic property $\phi$, the polynomial is reconstructed with respect to the center of the target cell, $(x_i, y_j)$. Figure \ref{fig:poly_uniform} illustrates the two-dimensional uniform grid and the stencil used to reconstruct. The stencil is constructed from the cell-averaged macroscopic properties $\overline{\phi}_{i,j}$, $\overline{\phi}_{i+1,j}$, $\overline{\phi}_{i,j+1}$, and $\overline{\phi}_{i+1,j+1}$, obtained from the four neighboring cells. The reconstructed polynomial is expressed as:
\begin{equation}
    \phi(x,y)=C_{00} + C_{01} (x-x_i) + C_{10} (y - y_j) + C_{11} (x-x_i)(y-y_j) .
\end{equation}
To minimize directional bias, independent polynomials are reconstructed in four quadrants around the cell center. When a boundary or surface is present, the reconstruction uses a stencil in the opposite direction. After reconstructing the polynomial, macroscopic properties are interpolated to each particle's position. These properties include $U_i$, $T_{\mathrm{trn}}$, $T_{\mathrm{rot}}$, $T_{\mathrm{vib}}$, and $\Pi_{ij}$, following the approach of Kim et al.\cite{Kim24-poly}

\begin{figure}[!t]%
\centering
\includegraphics[width=75mm,clip]{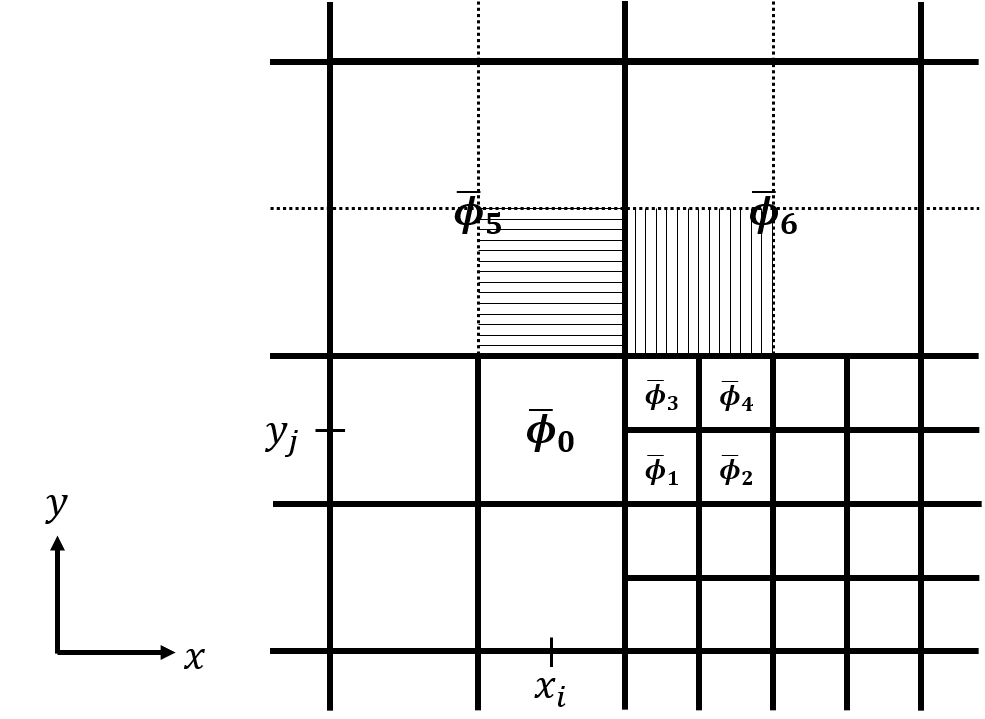}
\caption{A schematic of a non-uniform grid used for linear polynomial reconstruction.}%
  \label{fig:poly_nonuniform}%
\end{figure}

Handling non-uniform grids requires additional considerations. With a mesh refinement, neighboring cells may differ in size from the target cell. In such cases, macroscopic properties are rescaled to be consistent with the size of the target cell. Figure \ref{fig:poly_nonuniform} shows the two-dimensional non-uniform grid and the stencil. If a neighboring cell is larger than the target cell, its macroscopic property is evaluated by integrating the reconstructed polynomial over a subregion of the coarse cell. This subregion is chosen to match the size of the target cell, as illustrated by the shaded region in Fig. \ref{fig:poly_nonuniform}. If neighboring cells are smaller, their macroscopic properties are averaged and rescaled to match the size of the target cell. After rescaling, the reconstruction proceeds identically to the uniform grid case.

\section{\label{sec:implementation}Numerical implementation}

\subsection{\label{subsec:vhs}Viscosity and Mean Collision Time from the Chapman–Enskog Theory}


For the FPM model, the viscosity must be specified for particle evolution. According to Chapman–Enskog theory, the viscosity for the variable hard sphere (VHS) model is expressed as:\cite{Book-Bird94-DSMC}
\begin{equation}
    \mu = \mu_{\mathrm{ref}} \bigg( \frac{T}{T_{\mathrm{ref}}} \bigg)^{c_\omega} ,
\end{equation}
where the reference viscosity is defined as:
\begin{equation}
    \mu_{\mathrm{ref}} = \frac{15 \sqrt{mk_BT_{\mathrm{ref}}/\pi}}{2(5-2c_\omega)(7-2c_\omega)d_{\mathrm{ref}}^2} .
\end{equation}


In addition to the viscosity, the mean collision time is needed to determine the rotational and vibrational relaxation times, $\tau_{\mathrm{rot}}$ and $\tau_{\mathrm{vib}}$, respectively. 
In the variable hard sphere (VHS) model, the mean collision time is given by:\cite{Book-Bird94-DSMC}
\begin{equation}
    \tau_c= \frac{(5-2c_\omega)(7-2c_\omega)}{30} \frac{\mu}{p} .
\end{equation}

\subsection{\label{subsec:eucken}Prandtl Number Using the Eucken Formula}


For diatomic gases, a theoretical expression for the Prandtl number is given by the Eucken formula:\cite{Book-Boyd17-gas}
\begin{equation}
    \mathrm{Pr} = \frac{14+2\xi_{\mathrm{vib}}}{19+2\xi_{\mathrm{vib}}} ,
\end{equation}
where vibrational degrees of freedom is defined as:
\begin{equation}
    \xi_{\mathrm{vib}} = \frac{\Theta_{\mathrm{vib}}/T}{\exp{(\Theta_{\mathrm{vib}}/T)}-1} .
\end{equation}

\subsection{\label{subsec:conserve}Momentum and energy conservation scheme}


Since the particle FPM model employs a finite number of particles, exact conservation of momentum and energy within each cell cannot be guaranteed. To enforce momentum and energy conservation, the correction scheme proposed by Kim et al. is employed.\cite{Kim24-FPM} The expected energy relaxations are computed as 
\begin{multline}
    e_{\mathrm{trn}} (T_{\mathrm{trn}}^{n+1}) = e_{\mathrm{trn}}(T_{\mathrm{trn}}^n)
    - \gamma_{\mathrm{trn-rot}} (T_{\mathrm{trn}}^n - T_{\mathrm{rot}}^n) \\[0.5em]
    - \gamma_{\mathrm{trn-vib}} (T_{\mathrm{trn}}^n - T_{\mathrm{vib}}^n) ,
\end{multline}
\begin{multline}
    e_{\mathrm{rot}} (T_{\mathrm{rot}}^{n+1}) = e_{\mathrm{rot}}(T_{\mathrm{rot}}^n)
    + \gamma_{\mathrm{trn-rot}} (T_{\mathrm{trn}}^n - T_{\mathrm{rot}}^n) \\[0.5em]
    - \gamma_{\mathrm{rot-vib}} (T_{\mathrm{rot}}^n - T_{\mathrm{vib}}^n) ,
\end{multline}
\begin{multline}
    e_{\mathrm{vib}}(T_{\mathrm{vib}}^{n+1}) = e_{\mathrm{vib}}(T_{\mathrm{vib}}^n)
    + \gamma_{\mathrm{trn-vib}} (T_{\mathrm{trn}}^n - T_{\mathrm{vib}}^n) \\[0.5em]
    + \gamma_{\mathrm{rot-vib}} (T_{\mathrm{rot}}^n - T_{\mathrm{vib}}^n) .
\end{multline}


For the vibrational energy, a summation-based correction scheme is applied. The difference between the expected and actual vibrational energy levels is calculated as: 
\begin{equation}
    I_{\Delta} = \mathrm{int} \Bigg( \frac{e_{\mathrm{vib}}(T_{\mathrm{vib}}^{n+1}) - e_{\mathrm{vib}}(T_{\mathrm{vib}}^*)}
                      {R\Theta_{\mathrm{vib}}} + \mathrm{Unif}(0,1) \Bigg) .
\end{equation}
A particle is randomly selected within the cell. If $I_\Delta > 0$, the vibrational level of the selected particle is increased by 1 and $I_\Delta$ is decreased by 1. Conversely, if $I_\Delta < 0$, the vibrational level is decreased by 1 and $I_\Delta$ is increased by 1. This process is repeated until $I_\Delta = 0$.


For the rotational energy, a scaling-based correction scheme is used. In this scheme, the rotational energy of each particle is rescaled as:
\begin{equation}
    \varepsilon_{\mathrm{rot}}^{n+1}
    = \sqrt{\chi_{\mathrm{rot}}}
    \cdot\varepsilon_{\mathrm{rot}}^n ,
\end{equation}
\begin{equation}
    \chi_{\mathrm{rot}}
    = \frac{e_{\mathrm{rot}}(T_{\mathrm{rot}}^{n+1})}{e_{\mathrm{rot}}(T_{\mathrm{rot}}^*)}
\end{equation}


To conserve momentum and total energy, particle velocities are adjusted using the following scaling-based correction scheme:
\begin{equation}
    c_i^{n+1}
    = U_i^n
    + \sqrt{\chi_{\mathrm{trn}}}
    \cdot(c_i^* - U_i^*) ,
\end{equation}
\begin{multline}
    \chi_{\mathrm{trn}} =
    \Big(
        e_{\mathrm{trn}}(T_{\mathrm{trn}}^{n})
        + e_{\mathrm{rot}}(T_{\mathrm{rot}}^{n})
        + e_{\mathrm{vib}}(T_{\mathrm{vib}}^{n}) \\[0.5em]
        - e_{\mathrm{rot}}(T_{\mathrm{rot}}^{n+1})
        - e_{\mathrm{vib}}(T_{\mathrm{vib}}^{n+1})
    \Big) \, \Big/\, 
        e_{\mathrm{trn}}(T_{\mathrm{trn}}^*)
\end{multline}

\section{\label{sec:results}Results and discussion}


\begin{table}[!htp]
\centering
\caption{The VHS parameters of nitrogen.}\label{table:VHS}
\begin{tabular}{llll}\bhline{0.8pt}
$\mu_{\mathrm{ref}} \, [{\rm kg \cdot m^{-1} \cdot s^{-1}}]$ 
& $T_{\mathrm{ref}} \, [{\rm K}]$ 
& $d_{\mathrm{ref}} \, [{\rm m}]$ 
& $c_{\omega}$ \\ 
\hline
$1.658 \times 10^{-5}$ 
& $273.15$ 
& $4.17\times10^{-10}$ 
& $0.74$ \\ \bhline{0.8pt}
\end{tabular}
\end{table}%

Two test cases are numerically investigated: relaxation to thermal equilibrium in a homogeneous flow and hypersonic flow over a vertical plate. All cases use nitrogen, with a molecular mass of $4.65\times10^{-26} \; {\rm kg}$ and a characteristic vibrational temperature of $3371.0 \;{\rm K}$. The molecular properties are determined based on the VHS parameters, as summarized in Table \ref{table:VHS}. The FPM and USP-FPM models are implemented in the SPARTA, an open-source DSMC solver developed by Sandia National Laboratories.\cite{Plimpton19-SPARTA} To reproduce the Landau–Teller relaxation rate, the prohibiting double relaxation method is incorporated into the DSMC framework.\cite{Zhang13-prohibiting}

\subsection{\label{subsec:relax_temp}Relaxation to thermal equilibrium in a homogeneous flow}


Relaxation to thermal equilibrium in a homogeneous flow is investigated to compare the temporal accuracy of the FPM and USP-FPM models with a large time step. The initial number density and temperatures are specified as $n_0=10^{24} \,{\rm m^{-3}}$, $T_{\mathrm{trn},0}=12000 \,{\rm K}$, $T_{\mathrm{rot},0}=8000 \,{\rm K}$, and $T_{\mathrm{vib},0}=4000 \,{\rm K}$, respectively. The rotational and vibrational collision numbers are fixed at $Z_{\mathrm{rot}}=5$ and $Z_{\mathrm{vib}}=50$, respectively. The computational domain consists of a single cell with $\Delta x = 0.001 \,{\rm m}$, with one million computational particles used. The DSMC method with a time step of $10^{-11} \,{\rm s}$ serves as the reference solution. The FPM and USP-FPM models are compared with a larger time step of $10^{-9} \,{\rm s}$, which corresponds to the mean collision time.


\begin{figure}[!t]%
\centering
\includegraphics[width=80mm,clip]{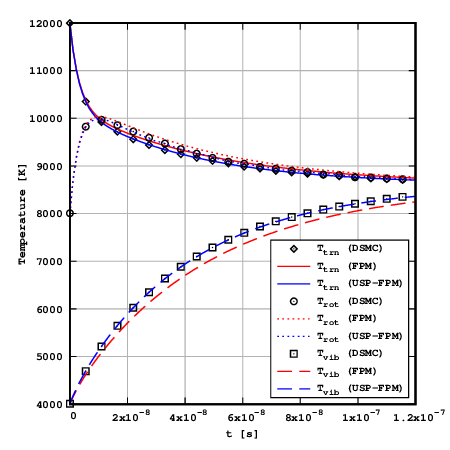}
\caption{Temporal evolution of translational, rotational, and vibrational temperatures during relaxation to thermal equilibrium in a homogeneous flow.}%
  \label{fig:homo_temp}%
\end{figure}%

Figure \ref{fig:homo_temp} shows the temporal evolution of translational, rotational, and vibrational temperatures toward equilibrium in a homogeneous flow. Because the rotational collision number is smaller than the vibrational collision number, translational and rotational temperatures reach equilibrium first. Subsequently, the vibrational mode relaxes more gradually toward equilibrium. The FPM model deviates from the DSMC results in all temperature modes, particularly underestimating the vibrational temperature. This discrepancy is attributed to the first-order temporal accuracy of the FPM model, which leads to inaccurate predictions of relaxation temperatures at large time steps. In contrast, the second-order USP-FPM model more accurately predicts the relaxation temperature, showing better agreement with the DSMC results.

\subsection{\label{subsec:couette}Hypersonic flow over a vertical plate}


\begin{table}[t]
\centering
\caption{Numerical parameters in the hypersonic flow over a vertical plate.}\label{table:plate_para}
\begin{small} 
\begin{tabular}{lllllll}\bhline{0.8pt}
Case
& \makecell{$\Delta x$ for \\ Level 1 \\ grid $[{\rm m}]$} 
& $\Delta t \;[{\rm s}]$
& ${\mathrm{N_{steady}}}$ 
& ${\mathrm{N_{sampling}}}$ 
& ${\mathrm{N_p}}$ \\ 
\hline
Fine  
& $0.0625$ 
& $1.6 \times 10^{-8}$ 
& $4 \times 10^5$ 
& $1 \times 10^5$ 
& $193 \;{\rm M}$  \\  
Coarse 
& $0.25$   
& $6.4 \times 10^{-8}$ 
& $1 \times 10^5$ 
& $1 \times 10^5$  
& $12 \;{\rm M}$  \\  
\bhline{0.8pt}
\end{tabular}
\end{small} 
\end{table}%

\begin{figure}[!t]%
\centering
\includegraphics[width=80mm,clip]{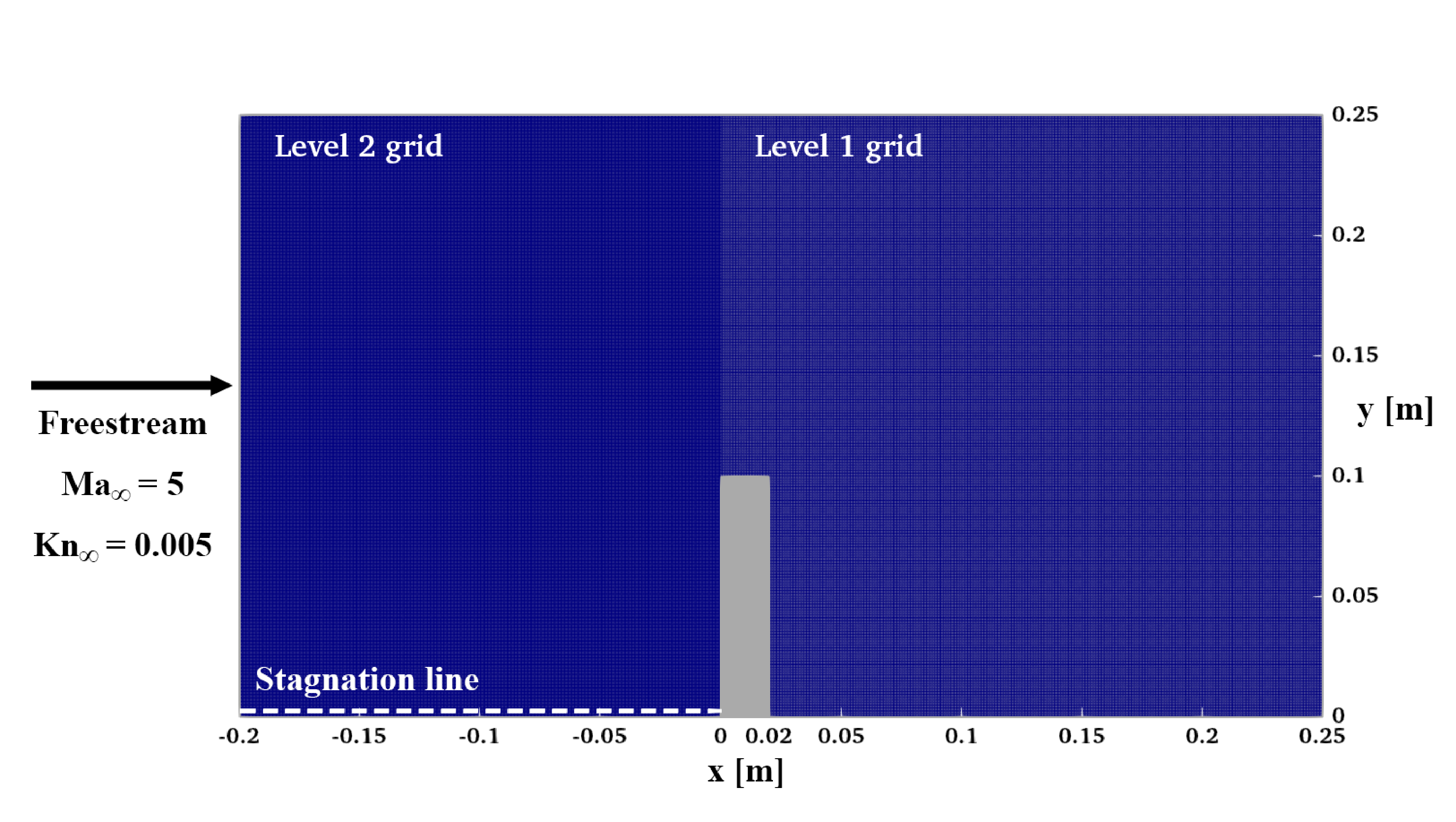}
\caption{Computational domain of the hypersonic flow over a vertical plate.}%
  \label{fig:homo_domain}%
\end{figure}%

To assess the accuracy of the USP-FPM model for predicting strong non-equilibrium effects, the hypersonic flow over a vertical plate is considered.\cite{Hepp20-CubicFP-diatomic, Kim24-FPM} The freestream number density and velocity are set to $n_\infty=1.3\times 10^{21} \,{\rm m^{-3}}$ and $U_\infty=2278.5 \,{\rm m / s}$, respectively. These conditions correspond to a Knudsen number of  $\mathrm{Kn}_{\infty} =0.005$ and a Mach number of $\mathrm{Ma}_\infty=5$. Fixed collision numbers of $Z_{\mathrm{rot}}=5$ and $Z_{\mathrm{vib}}=200$ are used. The vertical plate has a height of $0.2 \,{\rm m}$ and a width of $0.02 \,{\rm m}$. The plate is treated as a fully diffusive boundary with an isothermal wall temperature of $T_{\mathrm{wall}}=500 \,{\rm K}$. To reduce computational cost, a half-body simulation is performed by applying a reflective boundary along the x-axis. A two-level Cartesian grid is employed: Level 1 resolves the region behind the plate, and Level 2 resolves the region in front of the plate with half the cell size of Level 1. The computational domain is illustrated in Figure~\ref{fig:homo_domain}. The DSMC method with fine spatio-temporal resolution is used as a reference solution, where the Level 1 grid cell size and the time step are set to $\Delta x_{\mathrm{ref}}=0.0625 \; {\rm m}$ and $\Delta t_{\mathrm{ref}}=1.6\times 10^{-8} \,{\rm s}$, respectively. The results of the DSMC, FPM, and USP-FPM models are compared under coarse resolution with the Level 1 grid cell size of $4 \Delta x_{\mathrm{ref}}$ and the time step size of $4\Delta t_{\mathrm{ref}}$. For both resolutions, the number of simulation particles per cell is set to $100$ near the freestream inlet. Detailed numerical parameters are summarized in Table \ref{table:plate_para}.


\begin{figure}[!t]
  \centering
  \subfigure[Translational temperature profiles.]{%
    \includegraphics[width=0.45\textwidth]{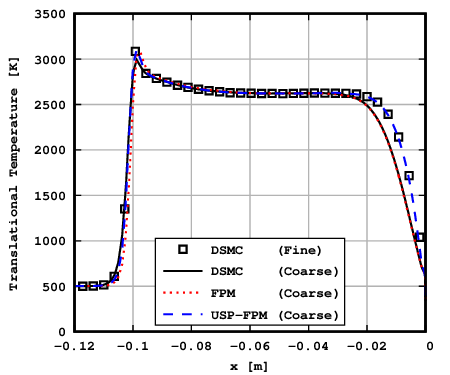}
    \label{fig:Ttrn}
  }
  \hfill
  \subfigure[Rotational temperature profiles.]{%
    \includegraphics[width=0.45\textwidth]{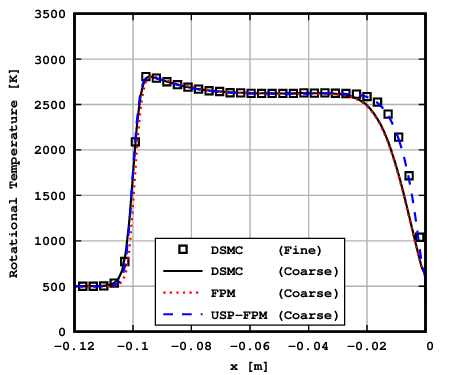}
    \label{fig:Trot}
  }
  \hfill
  \subfigure[Vibrational temperature profiles.]{%
    \includegraphics[width=0.45\textwidth]{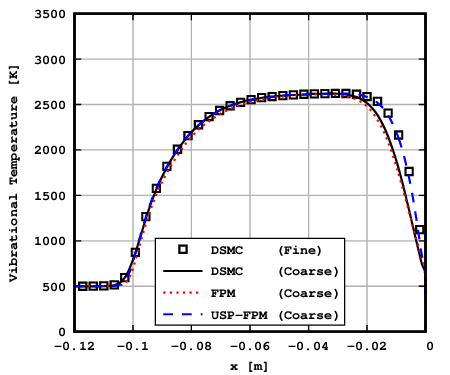}
    \label{fig:Tvib}
  }
  \caption{Translational, rotational, and vibrational temperature profiles along the stagnation line.}
  \label{fig:T_all}
\end{figure}

Figure \ref{fig:T_all} shows the translational, rotational, and vibrational temperature profiles along the stagnation line, as shown in Fig. \ref{fig:homo_domain}. A shock forms at $x=-0.1 \,{\rm m}$, initiating strong thermal non-equilibrium. The translational and rotational temperatures reach equilibrium at $x=-0.095 \,{\rm m}$, followed by the vibrational temperature at $x=-0.04 \,{\rm m}$. Beyond $x=-0.02 \,{\rm m}$, all temperatures decreases toward the wall temperature $T_{\mathrm{wall}}$. The DSMC method with coarse resolution underestimates the temperature near the plate compared to the fine-resolution DSMC results. A similar underestimation is observed in the FPM model at coarse resolution. In addition, the FPM model underestimates the post-shock temperature due to an inaccurate prediction of relaxation temperatures. In contrast, the USP-FPM model with coarse resolution shows good agreement with the DSMC method with fine resolution across all temperature modes.


\begin{table}[t]
\centering
\caption{Comparison of surface properties and total CPU time in the hypersonic flows over a vertical plate.}\label{table:CPU}
\begin{small} 
\begin{tabular}{llllll}\bhline{0.8pt}
\makecell{Numerical\\Method} 
& Case
& \makecell{Drag        \\$[{\rm N\cdot m^{-1}}]$   \\ (error [\%])} 
& \makecell{Peak heat flux\\$[{\rm kW\cdot m^{-2}}]$  \\ (error [\%])} 
& \makecell{Total CPU   \\ time $[{\rm h}]$ } \\ 
\hline
DSMC 
& Fine  
& $55.2  \;(-)$ 
& $31.9  \;(-)$ 
& $21916$  \\  
DSMC 
& Coarse 
& $55.2 \;(0.00)$  
& $39.9 \;(25.3)$  
& $504$  \\  
FPM 
& Coarse  
& $55.1 \;(0.13)$  
& $38.3 \;(20.0)$ 
& $884$  \\  
USP-FPM 
& Coarse  
& $55.1 \;(0.10)$ 
& $32.4 \;(1.71)$ 
& $2095$  \\  
\bhline{0.8pt}
\end{tabular}
\end{small} 
\end{table}%

Table \ref{table:CPU} presents the surface properties and total CPU time. All simulations were conducted using 192 cores on an AMD EPYC 9654 processor with a clock speed of $2.4 \, {\rm GHz}$. Compared to the DSMC method with fine resolution, both the DSMC method and the FPM model with coarse resolution accurately predict total drag with errors below $0.13 \,\%$, but exhibit peak heat flux errors exceeding $20.0 \,\%$. In contrast, the USP-FPM model with coarse resolution yields a drag error of $0.10 \,\%$ and a smaller peak heat flux error of $1.71 \,\%$. In terms of the computational efficiency, it achieves a $90.4 \,\%$ reduction in total CPU time while maintaining comparable accuracy in surface properties.

\section{\label{sec:conclusions}Conclusion}

This study proposes a USP-FPM model with second-order spatio-temporal accuracy. Second-order temporal accuracy is achieved by extending the USP-FP scheme to incorporate second-order energy relaxations, while spatial accuracy is obtained through a polynomial reconstruction method. The USP-FPM model is validated through two test cases: relaxation to thermal equilibrium in a homogeneous flow and hypersonic flow over a vertical plate. The results demonstrate that the USP-FPM model is more efficient than the DSMC method and the FPM model, due to its reduced sensitivity to variations in cell size and time step. Further study will extends the USP-FPM model to polyatomic gases and mixtures.


\section*{Acknowledgments}\label{Acknowledgments}

This work was supported by the National Research Foundation of Korea(NRF) grant funded by the Korea government(MSIT) (N01250638). This work was also supported by the National Supercomputing Center with supercomputing resources including technical support(KSC-2025-CRE-0427).



\begin{thebibliography}{99}
\bibitem{Book-Bird94-DSMC}
Bird, G. A.: \textit{Molecular Gas Dynamics and the Direct Simulation of Gas Flows}, Oxford University Press, Oxford, 1994.
\bibitem{KIM23-CompareCFDKinetic}
Kim, S., Park, W., and Jun, E.: A Comparative Study of CFD and Kinetic Models for Rarefied Gas Flows, \textit{J. Korean Soc. Aeronaut. Space Sci.}, \textbf{51} (2023), No. 12, pp. 849--859.
\bibitem{Jun11-Particle-conf}
Jun, E., Burt, J. M., and Boyd, I. D.: All-Particle Multiscale Computation of Hypersonic Rarefied Flow, \textit{AIP Conf. Proc.}, \textbf{1333} (2011), pp. 557--562.
\bibitem{Jun13-Particle-Conf}
Jun, E., Boyd, I., and Burt, J.: Assessment of an All-Particle Hybrid Method for Hypersonic Rarefied Flow, \textit{Proc. 51st AIAA Aerospace Sci. Meeting}, 2013.
\bibitem{Jun18-LD-DSMC}
Jun, E. and Boyd, I. D.: Assessment of the LD-DSMC Hybrid Method for Hypersonic Rarefied Flow, \textit{Comput. Fluids}, \textbf{166} (2018), pp. 123--138.
\bibitem{Jun18-FPDSMC-cylinder}
Jun, E., Gorji, M. H., Grabe, M., and Hannemann, K.: Assessment of the Cubic Fokker–Planck–DSMC Hybrid Method for Hypersonic Rarefied Flows Past a Cylinder, \textit{Comput. Fluids}, \textbf{168} (2018), pp. 1--13.
\bibitem{Jun19-FPDSMC-diatomic}
Jun, E.: Cubic Fokker–Planck–DSMC Hybrid Method for Diatomic Rarefied Gas Flow Through a Slit and an Orifice, \textit{Vacuum}, \textbf{159} (2019), pp. 125--133.
\bibitem{Kim24-FPDSMC}
Kim, S. and Jun, E.: An Evaluation of the Hybrid Fokker–Planck–DSMC Approach for High-Speed Rarefied Gas Flows, \textit{Comput. Fluids}, \textbf{285} (2024), p. 106456.
\bibitem{Yai23-SBGK}
Yao, S., Fei, F., Luan, P., Jun, E., and Zhang, J.: Extension of the Shakhov Bhatnagar–Gross–Krook Model for Nonequilibrium Gas Flows, \textit{Phys. Fluids}, \textbf{35} (2023), p. 037102.
\bibitem{Park24-BGK-PDF}
Park, W., Kim, S., Pfeiffer, M., and Jun, E.: Evaluation of Stochastic Particle Bhatnagar–Gross–Krook Methods with a Focus on Velocity Distribution Function, \textit{Phys. Fluids}, \textbf{36} (2024), p. 027113.
\bibitem{Jenny10_linearFP}
Jenny, P., Torrilhon, M., and Heinz, S.: A Solution Algorithm for the Fluid Dynamic Equations Based on a Stochastic Model for Molecular Motion, \textit{J. Comput. Phys}., \textbf{229} (2010), pp. 1077--1098.
\bibitem{Gorji11_cubicFP}
Gorji, M. H., Torrilhon, M., and Jenny, P.: Fokker–Planck Model for Computational Studies of Monatomic Rarefied Gas Flows, \textit{J. Fluid Mech.}, \textbf{680} (2011), pp. 574--601.
\bibitem{Mathiaud15_ESFP}
Mathiaud, J. and Mieussens, L.: A Fokker–Planck Model of the Boltzmann Equation with Correct Prandtl Number, \textit{J. Stat. Phys.}, \textbf{162} (2015), No. 2, pp. 397--414.
\bibitem{Gorji21_QEFP}
Gorji, M. H. and Torrilhon, M.: Entropic Fokker–Planck Kinetic Model, \textit{J. Comput. Phys.}, \textbf{430} (2021), p. 110034.
\bibitem{Kim22-FPNonlinear}
Kim, S. and Jun, E.: A Stochastic Particle Fokker–Planck Method with Nonlinear Production Terms for a Variable Hard-Sphere Gas, \textit{Phys. Fluids}, \textbf{34} (2022), p. 086111.
\bibitem{Jun19-CompareFP}
Jun, E., Pfeiffer, M., Mieussens, L., and Gorji, M. H.: Comparative Study Between Cubic and Ellipsoidal Fokker–Planck Kinetic Models, \textit{AIAA J.}, \textbf{57} (2019), pp. 2524--2533.
\bibitem{Kim24-AssessFP-Conf}
Kim, S. and Jun, E.: Assessment of Various Fokker–Planck Methods for Hypersonic Rarefied Flows, \textit{AIP Conf. Proc.}, (2024), p. 060006.
\bibitem{Gorji13_CubicFP-diatomic}
Gorji, M. H. and Jenny, P.: A Fokker–Planck Based Kinetic Model for Diatomic Rarefied Gas Flows, \textit{Phys. Fluids}, \textbf{25} (2013), No. 6, p. 062002.
\bibitem{Jun18-CubicFP-Slit}
Jun, E., Grabe, M., and Hannemann, K.: Cubic Fokker–Planck Method for Rarefied Monatomic Gas Flow Through a Slit and an Orifice, \textit{Comput. Fluids}, \textbf{175} (2018), pp. 199--213.
\bibitem{Mathiaud17-ESFP-polyatomic}
Mathiaud, J. and Mieussens, L.: A Fokker–Planck Model of the Boltzmann Equation with Correct Prandtl Number for Polyatomic Gases, \textit{J. Stat. Phys.}, \textbf{168} (2017), No. 5, pp. 1031--1055.
\bibitem{Hepp20-CubicFP-diatomic}
Hepp, C., Grabe, M., and Hannemann, K.: Master Equation Approach for Modeling Diatomic Gas Flows with a Kinetic Fokker–Planck Algorithm, \textit{J. Comput. Phys.}, \textbf{418} (2020), p. 109638.
\bibitem{Kim24-FPM}
Kim, S. and Jun, E.: A Stochastic Fokker–Planck–Master Model for Diatomic Rarefied Gas Flows, \textit{J. Comput. Phys.}, \textbf{506} (2024), p. 112940.
\bibitem{Gorji12-CubicFP-monamix}
Gorji, M. H. and Jenny, P.: A Kinetic Model for Gas Mixtures Based on a Fokker–Planck Equation, \textit{J. Phys.: Conf. Ser.}, \textbf{362} (2012), p. 012042.
\bibitem{Hepp20-CubicFP-HSmix}
Hepp, C., Grabe, M., and Hannemann, K.: A Kinetic Fokker–Planck Approach to Model Hard-Sphere Gas Mixtures, \textit{Phys. Fluids}, \textbf{32} (2020), No. 2, p. 027103.
\bibitem{Hepp20-CubicFP-VHSmix}
Hepp, C., Grabe, M., and Hannemann, K.: A Kinetic Fokker–Planck Approach for Modeling Variable Hard-Sphere Gas Mixtures, \textit{AIP Adv.}, \textbf{10} (2020), No. 8, p. 085219.
\bibitem{Kim25-ESFP-monamix}
Kim, S. and Jun, E.: A Particle Fokker–Planck Method for Rarefied Gas Flows of Monatomic Mixtures, \textit{Phys. Fluids}, \textbf{37} (2025), No. 1, p. 016104.
\bibitem{Kim25-ESFPM-dimix}
Kim, S. and Jun, E.: A Stochastic Particle Method Based on the Fokker–Planck Master Equation for Rarefied Gas Flows of Diatomic Mixtures, \textit{Phys. Fluids}, \textbf{37} (2025), No. 3, p. 036111.
\bibitem{Kim23-AssessmentFP}
Kim, S., Gorji, M. H., and Jun, E.: Critical Assessment of Various Particle Fokker–Planck Models for Monatomic Rarefied Gas Flows, \textit{Phys. Fluids}, \textbf{35} (2023), No. 4, p. 046117.
\bibitem{Kim24-USPFP}
Kim, S., Park, W., and Jun, E.: A Second-Order Particle Fokker–Planck Model for Rarefied Gas Flows, \textit{Comput. Phys. Commun.}, \textbf{304} (2024), p. 109323.
\bibitem{Cui25-MSP}
Cui, Z., Feng, K., Ma, Q., and Zhang, J.: A Multiscale Stochastic Particle Method Based on the Fokker–Planck Model for Nonequilibrium Gas Flows, \textit{J. Comput. Phys.}, \textbf{520} (2025), p. 113458.
\bibitem{Kim24-poly}
Kim, J., Kim, S., and Jun, E.: A Spatial Interpolation for a Stochastic Particle Fokker–Planck Model Using a Polynomial Reconstruction, \textit{Phys. Fluids}, \textbf{36} (2024), No. 12, p. 127171.
\bibitem{Gorji14-effect}
Gorji, M. H. and Jenny, P.: An Efficient Particle Fokker–Planck Algorithm for Rarefied Gas Flows, \textit{J. Comput. Phys.}, \textbf{262} (2014), pp. 325--343.
\bibitem{Fei22-USPBGK-poly}
Fei, F., Hu, Y., and Jenny, P.: A Unified Stochastic Particle Method Based on the Bhatnagar–Gross–Krook Model for Polyatomic Gases and Its Combination with DSMC, \textit{J. Comput. Phys.}, \textbf{471} (2022), p. 111640.
\bibitem{Mathiaud22-ESBGK}
Mathiaud, J., Mieussens, L., and Pfeiffer, M.: An ES-BGK Model for Diatomic Gases with Correct Relaxation Rates for Internal Energies, \textit{Eur. J. Mech. B Fluids}, \textbf{96} (2022), pp. 65--77
\bibitem{Gillespie00-tau-leap}
Gillespie, D. T.: The Chemical Langevin Equation, \textit{J. Chem. Phys.}, \textbf{113} (2000), No. 1, pp. 297--306.
\bibitem{Chatterjee05-modified-tau-leap}
Chatterjee, A., Vlachos, D. G., and Katsoulakis, M. A.: Binomial Distribution Based $\tau$-Leap Accelerated Stochastic Simulation, \textit{J. Chem. Phys.}, \textbf{122} (2005), No. 2.
\bibitem{Book-Boyd17-gas}
Boyd, I. D. and Schwartzentruber, T. E.: \textit{Nonequilibrium Gas Dynamics and Molecular Simulation}, Cambridge University Press, Cambridge, 2017.
\bibitem{Plimpton19-SPARTA}
Plimpton, S. J., Moore, S. G., Borner, A., Stagg, A. K., Koehler, T. P., Torczynski, J. R., and Gallis, M. A.: Direct Simulation Monte Carlo on Petaflop Supercomputers and Beyond, \textit{Phys. Fluids}, \textbf{31} (2019), No. 8, p. 086101.
\bibitem{Zhang13-prohibiting}
Zhang, C. and Schwartzentruber, T. E.: Inelastic Collision Selection Procedures for Direct Simulation Monte Carlo Calculations of Gas Mixtures, \textit{Phys. Fluids}, \textbf{25} (2013), No. 10, p. 106105.











\end{thebibliography}
\end{document}